\documentclass[iop]{emulateapj}
\usepackage{amsmath,amssymb,times,graphics}

\newcommand{\ha}{H\ensuremath{\alpha}}
\newcommand{\ewha}{EW\ensuremath{~({{\rm H}{\alpha}})}}
\newcommand{\lha}{\ensuremath{{\rm L}_{\rm H\alpha}/{\rm L}_{\rm bol}}}
\newcommand{\kms}{km~s\ensuremath{^{-1}}}

\newcommand{\rv}{\ensuremath{V_{r}}}
 
\newcommand{\masyr}{mas~yr\ensuremath{^{-1}}}

\newcommand{\Msun}{M\ensuremath{_{\sun}}}

\newcommand{\slowpokes}{SLoWPoKES}
\newcommand{\rz}{\ensuremath{r-z}}
\newcommand{\drv}{\ensuremath{\Delta{\rm V_{\rm r}}}}
\newcommand{\Pf}{P\ensuremath{_{\rm f}}}
\newcommand{\Nai}{\ion{Na}{1}}
\newcommand{\Cai}{\ion{Ca}{1}}

\defcitealias{Dhital2010}{Paper I}
\defcitealias{Lepine2007b}{LRS07}

\shorttitle{Refined Metallicity Indices for M Dwarfs}
\shortauthors{Dhital et al.}

\begin{document}

\journalinfo{{\it The Astronomical Journal}, accepted}
\submitted{Submitted: October 20, 2011; accepted: December 9, 2011; published:}

\title{Refined Metallicity Indices for M Dwarfs Using the
  {\slowpokes} Catalog of Wide, Low-mass Binaries}

\author{
  Saurav Dhital      \altaffilmark{1,2,7},
  Andrew A.\ West    \altaffilmark{2,8},
  Keivan G.\ Stassun \altaffilmark{1,3,4},
  John J.\ Bochanski \altaffilmark{5},
  Angela P. Massey      \altaffilmark{6,2},
  Fabienne A. Bastien   \altaffilmark{1}}
\altaffiltext{1}{Department of Physics \& Astronomy, Vanderbilt University,
  6301 Stevenson Center, Nashville, TN, 37235, USA; saurav.dhital@vanderbilt.edu} 
\altaffiltext{2}{Department of Astronomy, Boston University, 725
  Commonwealth Avenue, Boston, MA 02215, USA.}
\altaffiltext{3}{Department of Physics, Fisk University, 1000 17th Avenue N.,
  Nashville, TN 37208, USA.}
\altaffiltext{4}{MIT Kavli Institute for Astrophysics, 77
  Massachusetts Avenue, Cambridge, MA 02139, USA}
\altaffiltext{5}{Astronomy and Astrophysics Department, Pennsylvania
  State University, 525 Davey Laboratory, University Park, PA 16802, USA}
\altaffiltext{6}{Lusher Charter School, 5624 Freret St., New Orleans,
  LA 70115, USA}
\altaffiltext{7}{Visiting Astronomer, Kitt Peak National Observatory,
  National Optical Astronomy Observatory, which is operated by the
  Association of Universities for Research in Astronomy (AURA) under
  cooperative agreement with the National Science Foundation.}
\altaffiltext{8}{Visiting Investigator, Department of Terrestrial
  Magnetism, Carnegie Institute of Washington, 5241 Broad Branch Road,
  NW, Washington, DC 20015, USA.}

\begin{abstract}
We report the results from spectroscopic observations of 113 ultra-wide,
low-mass binary systems, composed largely of M0--M3 dwarfs, from the
{\slowpokes} catalog of common proper motion pairs identified in the
Sloan Digital Sky Survey. Radial velocities of each binary member were
used to confirm that they are co-moving and, consequently, to further
validate the high fidelity of the {\slowpokes} catalog. Ten stars
appear to be spectroscopic binaries based on broad or split spectral
features, supporting previous findings that wide binaries are likely
to be hierarchical systems. We measured the {\ha} equivalent width of the
stars in our sample and found that components of 81\% of the observed
pairs has similar {\ha} levels. The difference in {\ha} equivalent
width amongst components with similar masses was smaller than the range of
{\ha} variability for individual objects. We confirm that the
Lepine et al. $\zeta$-index traces iso-metallicity loci for most of
our sample of M dwarfs. However, we find a small systematic bias in
$\zeta$, especially in the early-type M dwarfs. We use our sample to
recalibrate the definition of $\zeta$. While representing a small
change in the definition, the new $\zeta$ is a significantly better
predictor of iso-metallicity for the higher mass M dwarfs. 
\end{abstract}


\section{Introduction}\label{Sec: intro}
Low-mass stars, generally defined as the regime bracketed by the
hydrogen-burning limit ($\sim$0.08~{\Msun}) and the onset of molecular
lines in the photosphere ($\sim$0.8~{\Msun}), make up $\sim$ 70\%
of the Milky Way's stars \citep{Bochanski2010} and are, perhaps, the
best tracers of the 
structure, dynamics, and evolutionary history of the Galaxy. However, their
intrinsic faintness has historically limited the construction of large
samples. In addition, the ubiquitous molecular features in their
photospheres and the resulting incomplete line lists has restricted
the accuracy and usefulness of theoretical atmospheric models. Large
 surveys, such as the Sloan Digital Sky Survey \citep[SDSS;][]{York2000}
and the Two Micron All Sky Survey \citep[2MASS;][]{Skrutskie2006},
have played a large role in advancing our understanding of low-mass
stars. With a photometric catalog of more than 33~million
\citep{Bochanski2010} and a spectroscopic catalog of more than 70,000
\citep{West2011} M dwarfs, SDSS has enabled studies of 
the spatial \citep{Bochanski2010} and kinematic distributions
\citep{Bochanski2007b, Fuchs2009} in the Milky Way; the 
mass and luminosity functions \citep{Covey2008, Bochanski2010}; and
magnetic activity \citep[e.g.,][]{West2008a, West2011, Kruse2010,
  Kowalski2009, Hilton2010} of low-mass stars.

The metallicity of low-mass stars remains an elusive parameter to
measure. Given the large number of M dwarfs in the Milky Way, an
absolute metallicity scale tied to an easily observable spectral index
would allow for the tracing of the formation history and the chemical
evolution of the Galaxy \citep[e.g.,][]{West2008a}, the dependence of
the fundamental mass--radius relation on metallicity at the bottom of
the main sequence \citep[e.g.,][]{Lopez-Morales2007}, and the
relationship between metallicity and the presence of planets
\citep[e.g.,][]{Laws2003, Valenti2008}. While spectral modeling has
allowed for metallicity determinations and well-defined metallicity
indices for warmer stars, such efforts in the late-K and M spectral
type regimes (e.g., \citealt*{Hauschildt1999a}; \citealt{Witte2011})
have met with notable problems due to the onset of broad molecular
lines at $\lesssim$4300~K and due to incomplete molecular line lists.
Some authors have tried to use photometric indices to infer the
metallicity \citep{Bonfils2005, Johnson2009, Schlaufman2010}, but
these techniques rely on trigonometric parallax measurements which are
uncommon for M dwarfs.

Some useful spectral features that correlate with metallicity
have been identified. In the near-infrared, \citet{Rojas-Ayala2010}
developed a metallicity indicator based on the strength of the {\Nai}
doublet, the {\Cai} triplet, and a temperature-sensitive water index.
This technique has so far only been calibrated over a limited range
but delivers the greatest precision among current techniques.
Meanwhile, much effort has gone into optical spectra. As the TiO band
in the optical spectrum becomes weaker with decreasing metallicity
\citep{Bessell1982}, the ratio of CaH and TiO molecular bands has been
used to distinguish M dwarfs from M subdwarfs \citep*{Kirkpatrick1991,
  Reid1995, Gizis1997, Lepine2003, Burgasser2006a}. Building on these
studies,  \citet*[][hereafter \citetalias{Lepine2007b}]{Lepine2007b}
defined the  metallicity-dependent quantity $\zeta$ using the
\citet{Reid1995} CaH2, CaH3, and TiO5 molecular band heads; this
allowed for the segregation of low-mass dwarfs into four classes:
dwarfs (dMs), subdwarfs (sdMs), extreme subdwarfs (esdMs), and ultra subdwarfs
(usdMs). These classes may also trace the Galactic populations to
which these stars belong: dMs were formed in the thin disk, sdMs in
the thick disk, and esdMs/usdMs in the halo. \citetalias{Lepine2007b}
calibrated the definition of $\zeta$ using the visual binary pairs
known at the time, including four sdM and two esdM
pairs. \citet*{Woolf2009} mapped the $\zeta$ index to an absolute
metallicity scale using dM binaries with a FGK companion of measurable 
absolute metallicity; but it suffers from significant scatter
($\sim$0.3~dex). 

Wide binary (or multiple) systems are ideal, coeval laboratories to
constrain and calibrate the observable properties of stars as the
components were presumably formed at the same time and from the same
primordial material but have evolved independently. In
\citet[][hereafter \citetalias{Dhital2010}]{Dhital2010} we identified
the Sloan Low-mass Wide Pairs of Kinematically Equivalent Stars
({\slowpokes}) catalog  consisting of 1342 ultra-wide, low-mass common
proper motion (CPM) binary systems from the SDSS Data Release 7
\citep[DR7;][]{Abazajian2009} by matching angular separations,
photometric distances, and proper motions.  The binary systems in the
catalog have at least one low-mass (spectral subtype K5 or later)
component, projected physical separations of $\sim$10$^3$--10$^5$~AU,
and distances of $\sim$50--800~pc. While most
{\slowpokes} pairs are disk dwarfs, 70 low-metallicity sdM and 21
white dwarf--dM pairs were identified based on their reduced
proper motions. A Galactic model---based on empirical constraints on
the stellar number density \citep{Bochanski2010, Juric2008} and
velocity \citep{Bochanski2007a} distributions in the Milky Way---was
used to assess the probability that the candidates were a chance
alignment of random stars; only pairs with such probabilities $\leq$
5\% were published in the {\slowpokes} catalog.  The overall fidelity
of the catalog is expected to be $\sim$98\%. Hence, the {\slowpokes}
catalog is a very clean and diverse source of CPM binary systems
to be used in follow-up studies. As the {\slowpokes} catalog spans a
wide range in mass  and a smaller, but still considerable, range in
metallicity, it is an ideal sample to constrain the $\zeta$ index as
well as to eventually map it to an absolute metallicity space.

Magnetic activity has been shown to decline with age, with activity
lifetimes of $\sim$1--2~Gyr for M0--M3 and $\sim$7--8~Gyr for M5--M7
dwarfs \citep{West2006, West2008a, West2011}. This monotonic decline
of activity with age is a signature of stellar spin-down and suggestive
of a gyrochronology-like age--rotation--activity relationship in M dwarfs
\citep{Skumanich1972, Barnes2003, Barnes2007, Delorme2010}. Leveraging
the coevality of components of the {\slowpokes} pairs is a good way of
testing this relationship. 

We have carried out a spectroscopic follow-up study of 113 CPM pairs
from the {\slowpokes} catalog. Section~\ref{Sec: observation} details
our observations and the data reduction procedures. In
Section~\ref{Sec: results} we use our radial velocities to assess the
fidelity of the observed {\slowpokes} pairs, use them to redefine the
\citetalias{Lepine2007b} $\zeta$ index, and examine the magnetic
activity properties of the {\slowpokes} pairs. The conclusions
are presented in Section~\ref{Sec: conclusions}.

\section{Observations \& Data Reduction}\label{Sec: observation}
The spectroscopic targets were selected from the {\slowpokes} catalog
based on their brightnesses, colors, and inferred mass ratios. Both
components were required to be brighter than $r\sim 17$ so as to
obtain the desired S/N within a reasonable integration time. Efforts
were made to obtain (i) an even distribution in {\rz} space for both
the primary and secondary components and (ii) a roughly equal number
of equal-mass (within 5\% of each other) and unequal-mass ($\gtrsim$
5\% of each other) systems. We estimated masses from
the  {\rz} colors based on \citet{Kraus2007}. 

Observations were carried out with the GoldCam spectrograph on the
KPNO 2.1m telescope on two separate observing runs on January 11--16,
2009 UT and March 26--31, 2010 UT. For both runs, the \#36 grating (1200
lines~mm$^{-1}$) in the first order, blazed at 7500~{\AA}, along with
the OG~550 order-blocking filter were used resulting in a wavelength
coverage of $\sim$6200--8200~{\AA} with a dispersion of
0.62~{\AA}~pixel$^{-1}$. A slit width of 2{\arcsec} was used to
maximize the number of photons collected yielding an effective
resolution of 1.8~{\AA} and a resolving power of 3500. Both components of a binary
were  observed at the same time by rotating the slit to align with the
position angle of the binary. While the rotation had to be done
manually and required $\sim$10~min of overhead time, it was more
efficient than observing each component separately.

Each night quartz flats and biases were taken before
the targets were observed; when the first half of the night was lost due to
weather, the flats and biases were taken in the morning. For
wavelength calibration, HeNeAr comparison arcs, along with the BG~38
order-blocking filter, were generally taken after each target or when
the CCD orientation was rotated. A suite of radial velocity standards
from \citet{Delfosse1998} were observed, which we used to assess our
radial velocity precision (see below). Similarly, a flux
standard---HZ~44, a bright sdO star---was observed each night during
the second run. Both  observing runs were conducted in bright time,
often during non-photometric seeing. A combination of clouds and high
winds caused the loss of 3--4 nights between the two runs.

\begin{figure}[tbh!]
  \begin{centering}
    \includegraphics[width=1\linewidth]{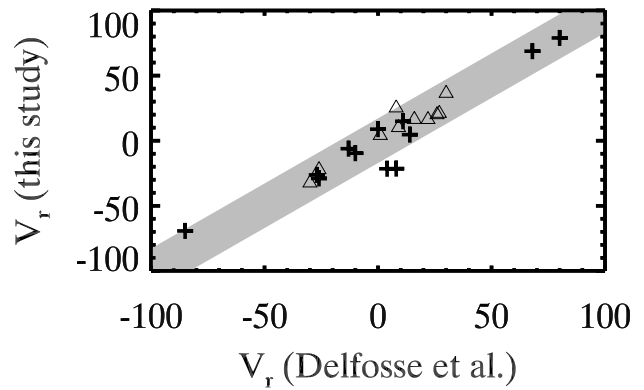}
    \caption{\protect\small
      Comparison of the radial velocities for the standard stars from
      the 2009 (triangles) and 2010 (pluses) observation
      runs selected from \citet{Delfosse1998} shows no systematic
      trends. The m.a.d. of the difference between the
      \citet{Delfosse1998} and the measured {\rv} was 5.7~{\kms}; we
      adopt this value as the error in our radial velocity measurements.
      The 3~m.a.d. regime is shaded in gray.
    }
    \label{Fig: rv_calib}
  \end{centering}
\end{figure}

\begin{center}
\begin{deluxetable}{lcc}
  \tablecaption{Radial Velocity Standards from \citep{Delfosse1998}}
  \tablecolumns{4}
  \tabletypesize{\scriptsize}
  \tablewidth{0pt}
  \tablehead{
    \colhead{Name} &
    \colhead{Spectral Type} &
    \colhead{{\rv} (\kms) }
}
\startdata
GJ 1057    &  M4  &   $\phn$27  \\
GJ 1093a   &  M4  &   $-$30   \\
GJ 1111    &  M8  &   $\phn$$\phn$9   \\
GJ 1156    &  M5  &   $\phn$$\phn$4   \\
Gl 70      &  M2  &   $-$26   \\
Gl 105b    &  M3  &   $\phn$26  \\
Gl 109     &  M2  &   $\phn$30  \\
G 165$-$08 &  M4  &   $\phn$$\phn$8   \\
Gl 205     &  M0  &   $\phn$$\phn$8  \\
Gl 251     &  M2  &   $\phn$22   \\
Gl 338     &  M0  &   $\phn$11   \\
Gl 380     &  K5  &   $-$26   \\
Gl 411     &  M2  &   $-$85   \\
Gl 412B    &  M5  &   $\phn$68   \\
Gl 450     &  M1  &   $\phn$$\phn$0   \\
Gl 514     &  M0  &   $\phn$14   \\
Gl 581     &  M2  &   $-$10   \\
Gl 623     &  M2  &   $-$27   \\
Gl 625     &  M1  &   $-$13   \\
LHS 1805   &  M4  &   $\phn$$\phn$1   \\
LHS 2520   &  M3  &   $\phn$80   \\
LHS 1885   &  M4  &   $\phn$16   \\
\enddata
\label{Tab: RVstd}
\end{deluxetable}
\end{center}
 
All spectra were bias-subtracted, flat-fielded, extracted, co-added,
wavelength-calibrated, and flux-calibrated\footnote{As
  flux-calibrations were not taken for the first run, we used a subset
  of radial velocity standards, which had absolute flux measurements
  as part of the Palomar-Michigan State Survey \citep{Reid1995}, as flux
  standards.} using standard IRAF\footnote{IRAF is distributed by the
  National Optical Astronomy Observatory, which is operated by the
  Association of Universities for Research in Astronomy (AURA) under
  cooperative agreement with the National Science Foundation.}
procedures, following the prescription detailed in
\citet{Massey1992}. Eighteen pairs where one of the components
(usually the fainter secondary) was not well-calibrated or had low S/N
were removed from the sample. The stars were then manually spectral
typed with the HAMMER pipeline \citep{Covey2007}; the error in
the process is expected to be smaller than one sub-type, as discussed
by \citet{West2011}.

The radial velocities ({\rv}) of the stars were measured by
cross-correlating the spectra using IDL routine {\sc xcorl.pro}
\citep{Mohanty2003, West2009} with the appropriate spectral type
templates from \citet{Bochanski2007b}, which are in the heliocentric 
rest frame. The cross-correlation was performed in the wavelength range
of 6600--7550~{\AA}. Since templates are only available for M0--L0 
dwarfs, M0 templates were used for K7 dwarfs. However, as each
spectral type spans a range in mass and temperature, the spectral
features in stars of the same subtype can differ significantly. 
This is probably the largest source of error in measuring {\rv} with
the cross-correlation techniques. However, cross-correlation with
templates typically yields {\rv} with $\lesssim$ 10~{\kms} precision
for SDSS spectra \citep{Bochanski2007a}  Other sources of error
include difference in resolution between the template and object
spectra and the accuracy of wavelength calibration, which was
$\lesssim$ 0.04~{\AA}/pixel rms (1.4~{\kms}) for all but a few of the objects.

To assess the errors in our radial velocities, we
cross-correlated the observed radial velocity standards
\citep{Delfosse1998} with appropriate templates from 
\citet{Bochanski2007b}. Figure~\ref{Fig: rv_calib} shows the
comparison between our measured values and the \citet{Delfosse1998}
values, which were measured from high-resolution spectra. Apart from
three outliers from the 2010 seasons, our values compare well with the
\citet{Delfosse1998}. The median absolute deviation (m.a.d.) of the
difference was 5.7~{\kms}; we adopt 6~{\kms} as the typical error in
our measurement of {\rv}.

The spectra were then corrected for the measured radial velocities to
be in the heliocentric rest frame and
fed back into the HAMMER pipeline to measure the equivalent width of
{\ha}; the molecular band strengths of CaH2, CaH3, and TiO features;
and the S/N of the spectra, which was measured in the region spanning
6500--6550~\AA. 

\section{Results}\label{Sec: results}

\begin{figure*}[t!]
  \begin{centering}
    \includegraphics[width=1\linewidth]{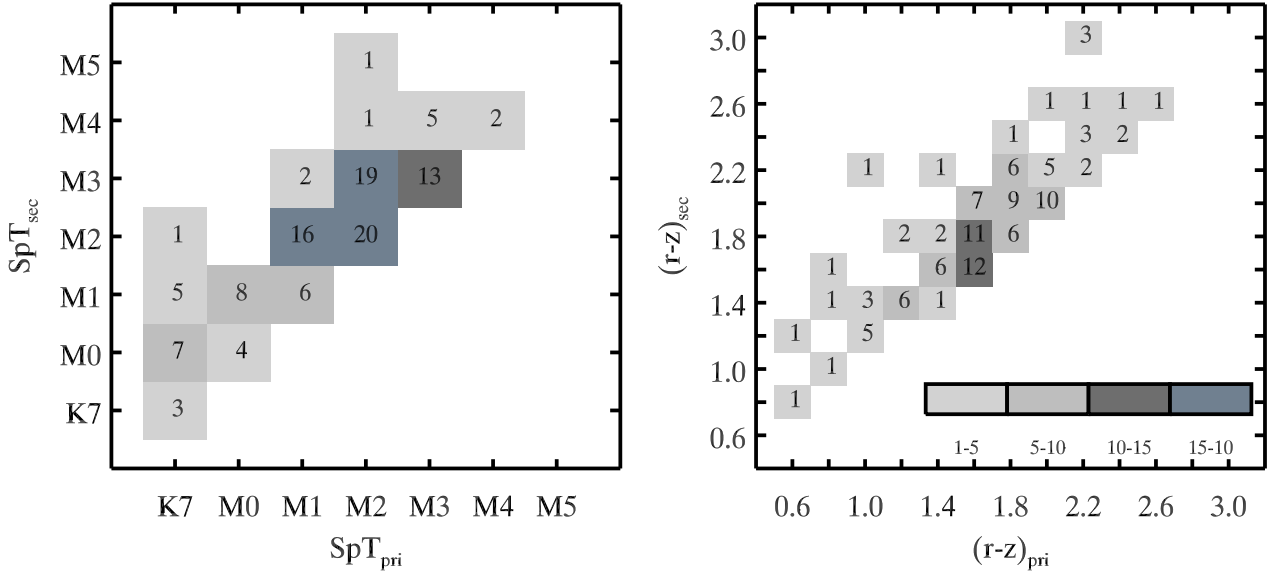}
    \caption{\protect\small
      The distribution of measured spectral types and {\rz} colors for both the
      primary and secondary components of the 113 {\slowpokes} pairs
      observed in our KPNO run. The spectral types are shown in 1
      subtype bins while the colors are in 0.2~mag bins. The
      background colors indicated the density in each bin, with the
      individual numbers printed as well. By definition, the primary
      always has a earlier spectral type and a bluer color than the secondary.}
    \label{Fig: spec_dist}
  \end{centering}
\end{figure*}

Figure~\ref{Fig: spec_dist} shows the spectral type and {\rz} color
distributions of the primary and the secondary components of the 113
{\slowpokes} pairs that were observed (Table~\ref{Tab: sample}); the number of pairs in
each bin is also shown. The observed sample, excluding pairs that were
rejected for low S/N or other reasons, spans the K7--M4 spectral
types ({\rz} $=$ 0.66--2.52) for the primary and K7--M5 ({\rz} $=$
0.77--3.08) for the secondary. Even though our observed sample was
limited to $r\sim$ 17 and, thus, a dearth of late-type M dwarfs was to
be expected, there are nonetheless 11 pairs with at least one
component later than M4 and only two pairs with both components later
than M4.

\begin{center}
\begin{deluxetable*}{cccccccrrcrrcrrcrr}
  \tablecaption{Properties of observed {\slowpokes} binaries}
  \tablecolumns{18}
  \tabletypesize{\scriptsize}
  \tablewidth{0pt}
  \tablehead{
    \colhead{ID} &
    \multicolumn{2}{c}{Right Ascension} &
    \colhead{} &
    \multicolumn{2}{c}{Declination} &
    \colhead{} &
    \multicolumn{2}{c}{distance} &
    \colhead{} &
    \multicolumn{2}{c}{$\mu_{\alpha}$} &
    \colhead{} &
    \multicolumn{2}{c}{$\mu_{\delta}$} &
    \colhead{} &
    \multicolumn{2}{c}{\rv} \\
    \colhead{}  & \colhead{A} & \colhead{B} &
    \colhead{}  & \colhead{A} & \colhead{B} &
    \colhead{}  & \colhead{A} & \colhead{B} &
    \colhead{}  & \colhead{A} & \colhead{B} &
    \colhead{}  & \colhead{A} & \colhead{B} &
    \colhead{}  & \colhead{A} & \colhead{B} \\
    \cline{2-3}
    \cline{5-6}
    \cline{8-9}
    \cline{11-12}
    \cline{14-15}
    \cline{17-18}
    \colhead{SLW J2000} &
    \multicolumn{2}{c}{(hh mm ss.s)} &
    \colhead{} &
    \multicolumn{2}{c}{(dd mm ss.ss)} &
    \colhead{} &
    \multicolumn{2}{c}{(pc)} &
    \colhead{} &
    \multicolumn{2}{c}{(\masyr)} &
    \colhead{} &
    \multicolumn{2}{c}{(\masyr)} &
    \colhead{} &
    \multicolumn{2}{c}{(\kms)}}
\startdata
1512+20 &  15 12 22.52 &  15 12 25.41 &&  +20 28 20.6 &  +20 28 12.3 &&   63 &   58 &&  -49 &  -48 &&    4 &    7 &&  -84.8 &  -77.8  \\
0831+36 &  08 31 23.12 &  08 31 23.16 &&  +36 54 41.8 &  +36 54 17.2 &&   70 &   83 &&   49 &   46 &&  -54 &  -57 &&  -30.8 &  -25.9  \\
0741+19 &  07 41 55.34 &  07 41 57.06 &&  +19 55 45.8 &  +19 55 33.3 &&   66 &   78 &&  -36 &  -35 &&  -27 &  -27 &&   31.2 &   47.8  \\
0957+37 &  09 57 57.18 &  09 57 55.63 &&  +37 56 02.4 &  +37 55 52.8 &&   87 &   75 &&  -23 &  -23 &&  -60 &  -59 &&  -23.6 &  -31.2  \\
1120+20 &  11 20 03.38 &  11 20 05.26 &&  +20 46 53.2 &  +20 46 54.9 &&   96 &  101 &&  -37 &  -41 &&   -2 &    0 &&   -7.6 &  -40.6  \\
0858+09 &  08 58 57.80 &  08 58 54.73 &&  +09 36 59.1 &  +09 37 23.7 &&   65 &   63 && -111 & -107 &&    6 &    6 &&   23.5 &   32.7  \\
1527+49 &  15 27 52.04 &  15 27 50.57 &&  +49 08 54.2 &  +49 09 47.4 &&   70 &   65 &&  -60 &  -63 &&   50 &   53 &&  -89.7 &  -90.2  \\
0734+28 &  07 34 50.75 &  07 34 49.23 &&  +28 17 39.7 &  +28 18 15.8 &&   62 &   75 &&  -27 &  -29 &&  -28 &  -29 &&   -2.3 &  -36.7  \\
1318+47 &  13 18 15.49 &  13 18 15.00 &&  +47 30 29.4 &  +47 31 33.7 &&   47 &   47 && -104 & -103 &&   33 &   35 &&  -60.1 &  -71.6  \\
1508+06 &  15 08 44.07 &  15 08 43.72 &&  +06 46 25.9 &  +06 46 35.5 &&  110 &  108 &&  -42 &  -44 &&    0 &    0 &&  -80.4 &  -51.3  \\
\enddata
\end{deluxetable*}
\end{center}

\begin{center}
\begin{deluxetable*}{cccccccrrcrrcrrcrrcrr}
  \tablecolumns{21}
  \tabletypesize{\scriptsize}
  \tablewidth{0pt}
  \tablecaption{}
  \tablenum{2}
  \tablehead{
    \colhead{ID} &
    \multicolumn{2}{c}{$r$} &
    \colhead{} &
    \multicolumn{2}{c}{\rz} &
    \colhead{} &
    \multicolumn{2}{c}{Spectral Type} &
    \colhead{} &
    \multicolumn{2}{c}{\ewha} &
    \colhead{} &
    \multicolumn{2}{c}{CaH2} &
    \colhead{} &
    \multicolumn{2}{c}{CaH3} &
    \colhead{} &
    \multicolumn{2}{c}{TiO5}\\
    \colhead{}  & \colhead{A} & \colhead{B} &
    \colhead{}  & \colhead{A} & \colhead{B} &
    \colhead{}  & \colhead{A} & \colhead{B} &
    \colhead{}  & \colhead{A} & \colhead{B} &
    \colhead{}  & \colhead{A} & \colhead{B} &
    \colhead{}  & \colhead{A} & \colhead{B} &
    \colhead{}  & \colhead{A} & \colhead{B} \\
    \cline{2-3}
    \cline{5-6}
    \cline{8-9}
    \cline{11-12}
    \cline{14-15}
    \cline{17-18}
    \cline{20-21}
    \colhead{SLW J2000}  &
    \multicolumn{2}{c}{(mag)} &
    \colhead{} &
    \multicolumn{2}{c}{} &
    \colhead{} &
    \multicolumn{2}{c}{} &
    \colhead{} &
    \multicolumn{2}{c}{(\AA)} &
    \colhead{} &
    \multicolumn{2}{c}{} &
    \colhead{} &
    \multicolumn{2}{c}{} &
    \colhead{} &
    \multicolumn{2}{c}{}}
\startdata
1512+20  &  15.01 &  16.87 &&   2.02 &   2.56 && M3 & M4 && -0.18 &  0.00 &&  0.45 &  0.36 &&  0.72 &  0.63 &&  0.44 &  0.34  \\
0831+36  &  16.13 &  16.51 &&   2.12 &   2.36 && M3 & M4 &&  2.36 &  7.93 &&  0.42 &  0.38 &&  0.69 &  0.62 &&  0.43 &  0.37  \\
0741+19  &  16.29 &  16.67 &&   2.22 &   2.46 && M3 & M4 &&  3.06 &  7.18 &&  0.43 &  0.41 &&  0.69 &  0.67 &&  0.42 &  0.38  \\
0957+37  &  17.28 &  17.34 &&   2.45 &   2.57 && M3 & M4 &&  4.11 &  7.16 &&  0.39 &  0.39 &&  0.65 &  0.64 &&  0.38 &  0.34  \\
1120+20  &  16.62 &  17.09 &&   2.13 &   2.32 && M3 & M3 &&  7.16 &  0.99 &&  0.46 &  0.44 &&  0.72 &  0.70 &&  0.45 &  0.45  \\
0858+09  &  15.58 &  15.76 &&   2.12 &   2.20 && M3 & M3 &&  5.12 &  2.57 &&  0.42 &  0.42 &&  0.65 &  0.67 &&  0.39 &  0.43  \\
1527+49  &  16.00 &  17.01 &&   2.19 &   2.57 && M3 & M4 &&  4.83 &  1.08 &&  0.32 &  0.41 &&  0.59 &  0.68 &&  0.31 &  0.41  \\
0734+28  &  14.40 &  14.78 &&   1.65 &   1.90 && M2 & M2 &&  4.64 &  0.00 &&  0.46 &  0.53 &&  0.69 &  0.76 &&  0.49 &  0.55  \\
1318+47  &  15.19 &  17.40 &&   2.20 &   2.95 && M2 & M5 &&  1.95 &  3.94 &&  0.41 &  0.33 &&  0.68 &  0.62 &&  0.42 &  0.27  \\
1508+06  &  15.66 &  17.04 &&   1.78 &   2.23 && M2 & M2 &&  3.53 &  0.64 &&  0.42 &  0.49 &&  0.67 &  0.75 &&  0.43 &  0.52  \\
\enddata
\tablecomments{The first 10 pairs are listed here; the full version of
  the table is available online.}
\label{Tab: sample}
\end{deluxetable*}
\end{center}

\subsection{Spectroscopic Binaries}\label{Sec: SBs}
\begin{figure*}[tbh!]
  \begin{centering}
    \includegraphics[width=1\linewidth]{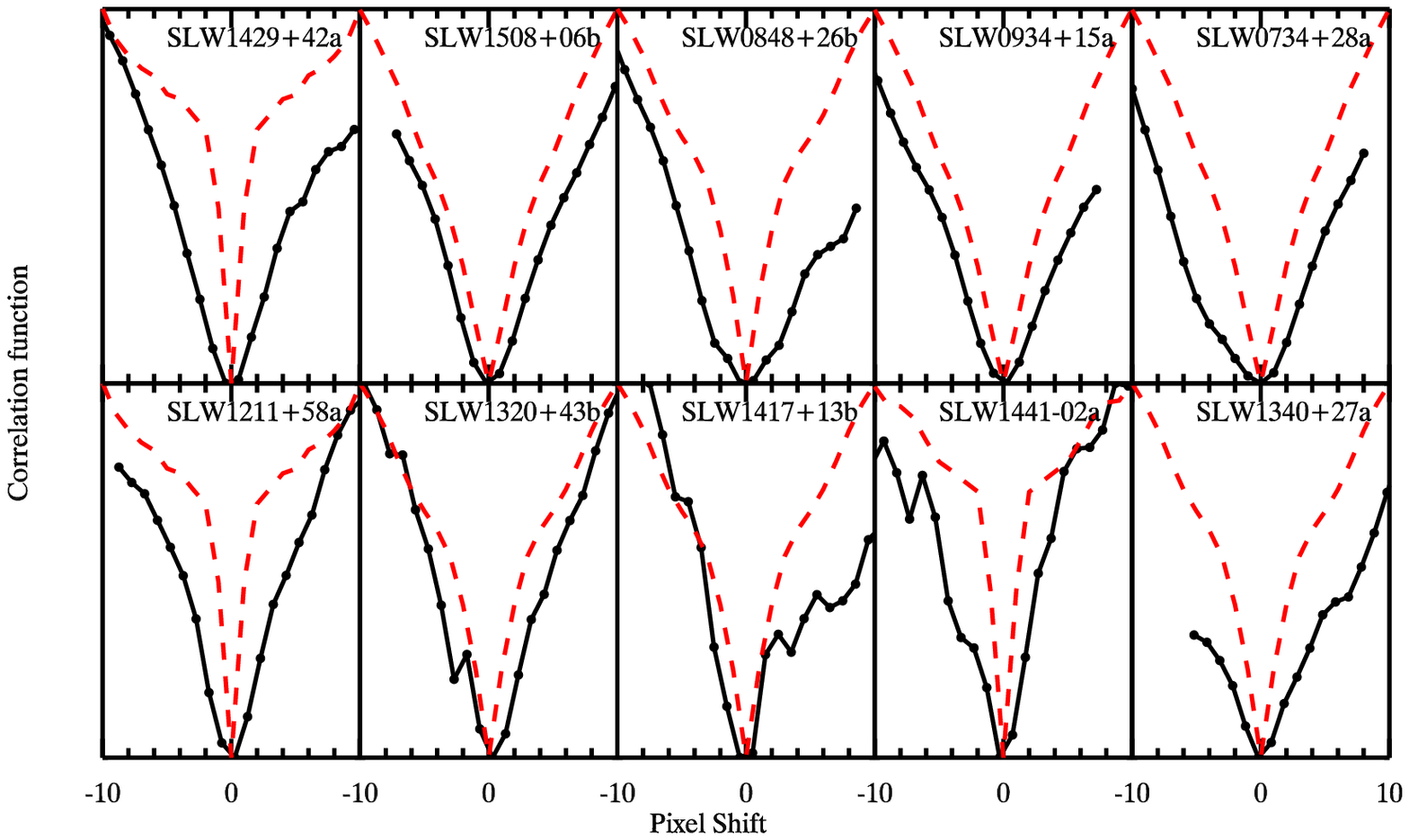}
    \caption{\protect\small
      Broad cross-correlation functions (solid lines) and/or split
      spectral features indicate the presence of a tight companion in
      the ten of the observed systems.  For reference, the auto-correlation function
      of our radial velocity standards are also shown (dashed
      lines). High-resolution spectra are needed to confirm the
      spectroscopic binaries. All spectra were corrected to the heliocentric rest frame.}
    \label{Fig: SBs}
  \end{centering}
\end{figure*}

As discussed in Section~\ref{Sec: observation}, the radial velocities
were measured by cross-correlating the program spectra with the
appropriate SDSS template spectra \citep{Bochanski2007b}. The
cross-correlation function (CCF) is used to determine the best match
between the templates and program spectra. Presence of multiple
turning points in the CCF as well as unusual broadening of the
spectrum can indicate the presence of a spectroscopic binary
\citep[SB;][]{Matijevic2010}. While such a detection is unambiguous
only in high-resolution spectra, we found possible evidence of SBs in
our low-resolution spectra.  Alternatively, the wide CCF could
correspond to fast rotators, although our $v~\sin~i$ resolution of 35~{\kms}
means they would have to be rotating at very high speeds.
Figure~\ref{Fig: SBs} shows the CCF for
the ten SB candidates (Table~\ref{Tab: SB}).  All ten candidates have
a relatively high S/N ratio, so the CCF is not a product of noisy
spectral features. For comparison, the CCF for the radial velocity
standards, which are presumably single stars, of the corresponding
spectral types are shown in red, dashed lines. High-resolution spectra
are required to confirm these SBs.

Previous studies have found that components of wide binaries are more
likely to have a companion as compared to single field stars. This
enhanced binarity has been ascribed to the ease of transfer of angular
momentum that facilitates the formation of close pairs \citep{Tokovinin1997,
  Bate2002b, Burgasser2005, Connelley2009} and/or the stability of
wide pairs in the field \citep{Law2010}. Among very low-mass wide
binaries, the frequency of tight companions is (50$\pm$11)\%
\citep{Faherty2010}. In a sample of nearby {\slowpokes} pairs,
\citet{Law2010} found that the bias-corrected higher-order
multiplicity was 45$^{+18}_{-16}$\%.  While only 10 of 113 pairs
(8.8\%) in this study have been identified as hierarchical, we were
probing a different kind of hierarchical systems than those found by
\citealt{Law2010}. Here we probed the extremely close pairs
(spectroscopic binaries) whereas \citealt{Law2010} probed systems with
separations larger than 8--10~AU. In fact, our results are consistent with the
\citealt{Law2010} findings but limited to the extremely close pairs.

\begin{center}
\begin{deluxetable}{ccl}
  \tablecaption{Candidate Spectroscopic Binaries}
  \tablecolumns{3}
  \tabletypesize{\scriptsize}
  \tablewidth{0pt}
  \tablehead{
    \colhead{Name} &
    \colhead{Spectral Type} &
    \colhead{Note}
}
\startdata
SLW 1211+58a   &   M0  &  \nodata \\
SLW 1320+43b   &   M1  &  wide CCF \\
SLW 1417+13b   &   M1  &  wide CCF \\
SLW 1441-02a   &   K7  &  \nodata  \\
SLW 1340+27a   &   M1  &  wide lines; wide CCF \\
SLW 1429+42a   &   M0  &  double peak \\
SLW 1508+06b   &   M2  &  wide CCF \\
SLW 0848+26b   &   M1  &  wide CCF \\
SLW 0934+15a   &   M2  &  line splitting; double peak \\
SLW 0734+28a   &   M   &  line splitting, wide CCF \\
\enddata
\label{Tab: SB}
\end{deluxetable}
\end{center}

\subsection{Fidelity of {\slowpokes} Pairs}\label{Sec: RV}
The observed pairs were identified in \citetalias{Dhital2010} based on a
matching of their position, distance, and proper motions. The third
velocity component, {\rv}, can be used to test the fidelity of the
observed pairs and, by extrapolation, of the {\slowpokes} catalog.  

\begin{figure*}[tbh!]
  \begin{centering}
    \includegraphics[width=0.47\linewidth]{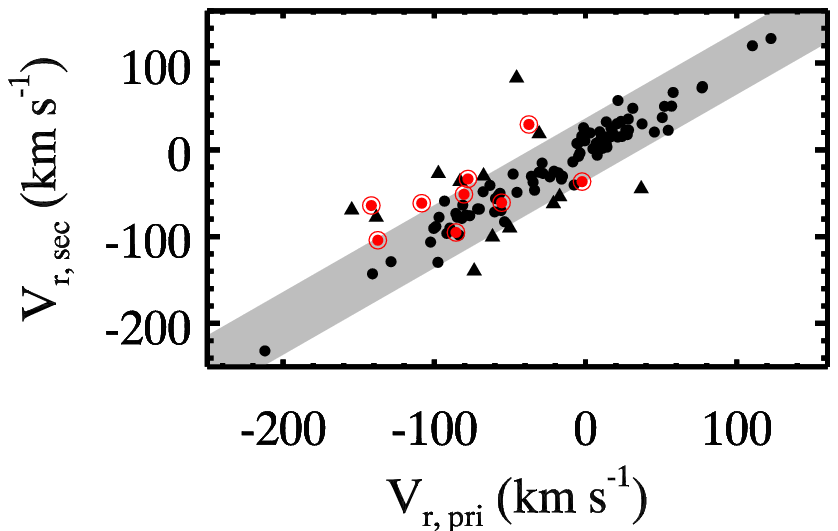}
    \includegraphics[width=0.47\linewidth]{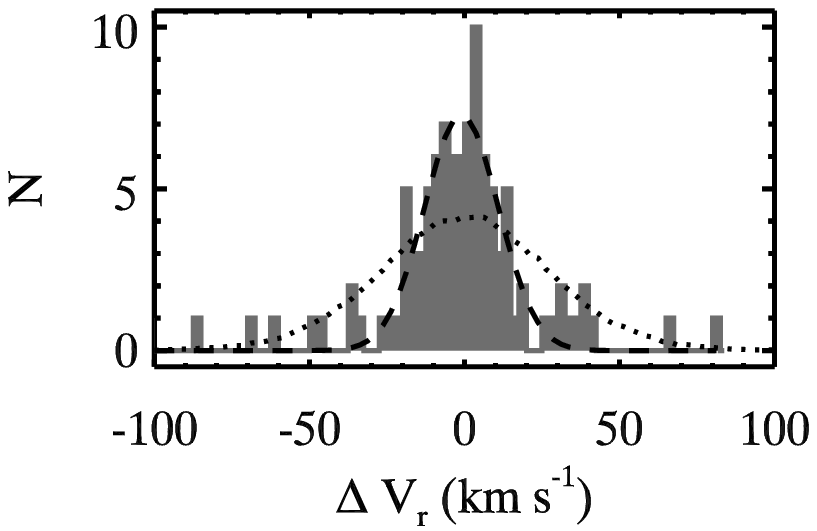}
    \caption{\protect\small
      {\it Left}: The radial velocity of the primary and the secondary
      components: the pairs with {\drv} $>3\sigma$ are shown as
      triangles while the candidate spectroscopic binaries, shown as red
      concentric circles, have larger {\drv}.
      {\it Right}: The distribution of difference in radial velocities
      between components of {\slowpokes} pairs observed in this
      program, with the Gaussian fit shown in dashed lines. 85\% of
      the sample has {\drv} $\le$ 3$\sigma$. All pairs with
      {\drv}$>$50~{\kms} have relatively low S/N.. The expected
      intrinsic scatter in {\drv} for two unassociated stars in the same 3-d
      position in the Galaxy and have matching proper motions, as
      calculated using the Galactic model from \citetalias{Dhital2010} is
      shown with dotted lines; it is a much larger dispersion compared
      to our observed sample of CPM pairs.}
    \label{Fig: rv}
  \end{centering}
\end{figure*}

Figure~\ref{Fig: rv} shows the radial velocities of the primary
component against that of the secondary in the left panel and the
distribution of their differences in the right panel. The identified
candidate SBs are shown as concentric circles.  Excluding the
ten candidate SBs, 90 of the remaining 103 pairs (i.e., 87.4\%) have
{\drv} within 3~$\sigma$ of the mean; the 3~$\sigma$ region is shown in
gray in the left panel. Overall, the {\drv} distribution is well-fit
by a  Gaussian with $\mu=-0.97 \pm 0.80$~{\kms} and $\sigma= 12.04 \pm
0.80$~{\kms}, shown as the dashed line though there are more outliers
than expected (see below). As $\sigma \approx $1.4826 m.a.d for large normally
distributed populations, $\sigma_{\drv}$ ( $= \sqrt{2}\times
  \sigma_{\rv} = \sqrt{2}\times1.4826\times$ 5.7~{\kms} $=$
  11.95~{\kms}) is in excellent agreement with the m.a.d. we measured 
for our radial velocity standards.

We checked that the {\drv} distribution of our binaries is indeed
distinct from physically unassociated stars. We used the Galactic model
from \citetalias{Dhital2010} that gives the expected 3-d velocity
distribution for any position in the Galaxy or, if desired, a randomly
chosen velocity from that distribution. We compared with a sample that
has been selected in a similar manner to the pairs in our observed
sample. Hence, at the Galactic positions of each of the 
observed pairs, we generated pairs of 3-d velocities until a pair with
matching proper motions was found. The proper motion matching criteria
was the same as that in \citetalias{Dhital2010}. For statistical
robustness, we conducted 10$^6$ realizations of this simulation; the
normalized histogram of the resultant distribution is shown in dotted
lines in the right panel of Figure~\ref{Fig: rv}. For a quantitative
assessment of the difference between the simulated and observed {\drv}
distributions, we performed the Kolmogorov-Smirnov test
\citep{Press1992} and found a 0.93\% probability that the two were
drawn from the same parent population. We conclude the {\drv}
distribution of our observed binaries is much narrower than the
scatter expected of two unassociated stars.

\begin{figure*}[tbh!]
  \begin{centering}
    \includegraphics[width=0.47\linewidth]{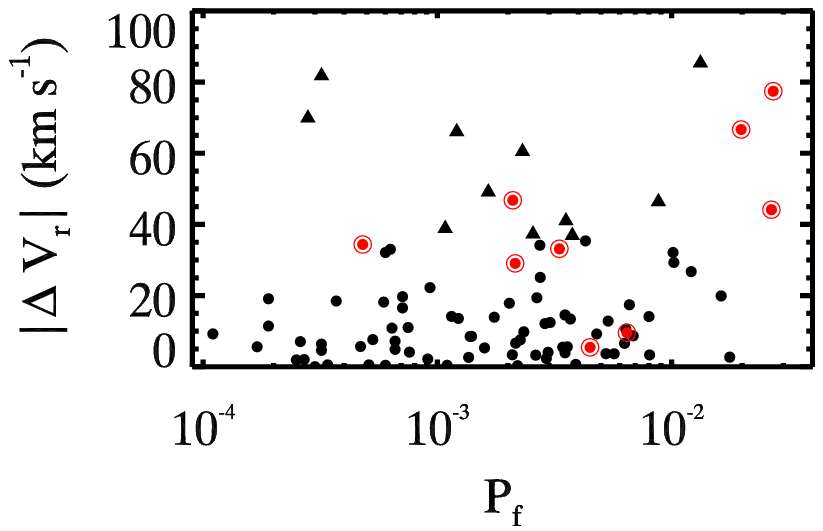}
    \includegraphics[width=0.47\linewidth]{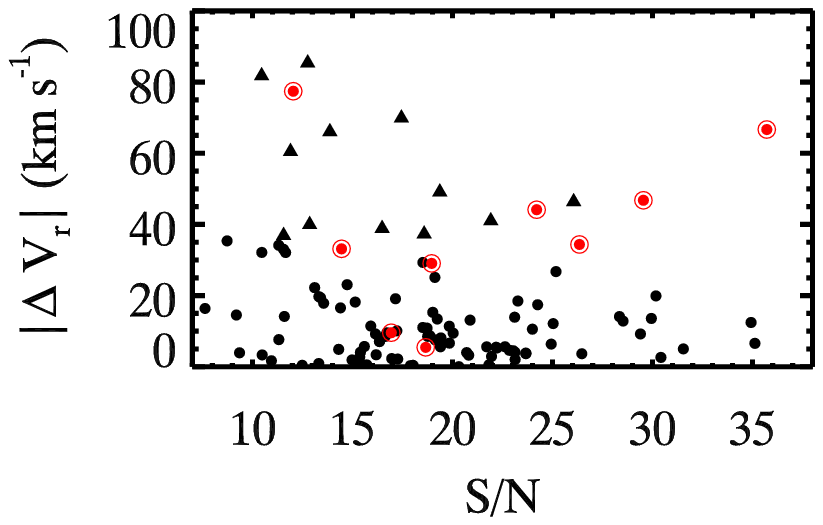}
    \caption{\protect\small
      {\it Left:} The probability of chance alignment ({\Pf}) calculated using the
      Galactic model by \citetalias{Dhital2010} vs. the difference in
      radial velocity between components for 113 {\slowpokes} CPM
      pairs observed in this program. 
      {\it Right}: Systems with a lower S/N (red) have a higher
      dispersion in {\drv}, suggesting the observed discrepancy might be due
      to the lower S/N. Candidate SBs are shown as red, concentric
      circles while pairs with {\drv} $> 3\sigma$ are shown as triangles.
    }
    \label{Fig: rv.pf}
  \end{centering}
\end{figure*}

Thirteen (12.6\%) of the pairs have {\rv} that disagree at
$>$3$\sigma$.  The {\slowpokes} catalog only contains pairs with
probability of chance alignment, {\Pf}, less than 5\%, meaning fewer
than five of 103 pairs were expected to be false positives. In fact,
{\Pf} was tabulated for each pair in \citetalias{Dhital2010}; the
cumulative sum of chance alignments was only 0.3\%, implying $<$1 pair
was expected to be false positives. As shown in the left panel of
Figure~\ref{Fig: rv.pf}, there is no trend in {\drv} as a function of
the {\Pf}.  There are discrepant pairs at all values of
{\Pf}. However, as can be seen in the right panel of Figure~\ref{Fig:
  rv.pf}, there is a significant trend of {\drv} with the 
S/N in the spectra. The pairs with the largest {\drv} values are
at low S/N while there are no discrepant pairs at high S/N.  This
suggests that the cross-correlation process and the measurements of
{\rv}  might have been adversely affected by the noise, yielding
noisier radial velocities. In addition, given the large higher-order multiplicity
fraction seen in {\slowpokes} \citep[45\%;][]{Law2010}, the presence
of more SBs in our observed sample cannot be ruled out by our low-resolution
spectra.

In summary, the vast majority of the sample pairs show agreement in
their radial velocities, as expected for physical binaries. Pairs with discrepant
{\rv}s have spectra with low observed S/N or are (candidate) hierarchical
systems with a spectroscopic binary.

\subsection{Metallicity Index Calibration}\label{Sec: met}

\begin{figure*}[tbh!]
  \begin{centering}
    \includegraphics[width=1\linewidth]{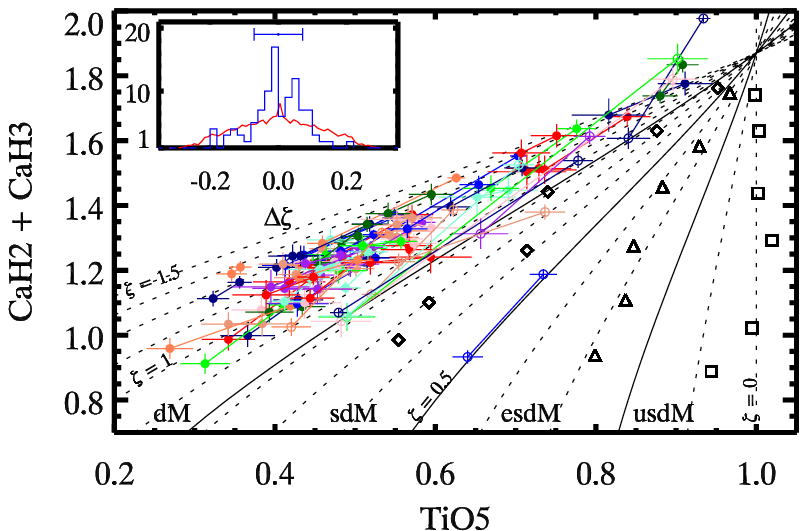}
    \caption{\protect\small
      Distribution of the observed M dwarf binaries, with components
      of a system connected by solid lines, in the CaH/TiO space confirms
      the iso-$\zeta$ contours with $\zeta=$ 0.0, 0.1, 0.2, ..., 1.5
      (dotted lines), albeit with some deviation at the highest
      values (towards top right). The distribution of $\Delta\zeta$
      (blue histogram; inset) peaks at $\sim$0 and is significantly
      different from a distribution for two randomly paired dMs (red
      solid line). However, the scatter in $\Delta\zeta$ is large; the 
      bar at the middle-top of the inset shows the scatter in the
      error of $\Delta\zeta$ ($\sigma_{\Delta\zeta}$). 
      The solid contours demarcate the boundary between the
      dM/sdM/esdM/usdM classes while the K7--M5 spectral standards for
      the sdM (diamonds), esdM (triangles), and usdM (squares) classes
      are shown \citepalias{Lepine2007b}. Early-type M stars are at
      the upper right, late-types at the lower left.}   
    \label{Fig: met0}
  \end{centering}
\end{figure*}

Figure~\ref{Fig: met0} shows the observed {\slowpokes} pairs, with the
components of each pair connected with a solid line, in the CaH--TiO
space with the dotted lines showing the iso-$\zeta$ lines from
\citetalias{Lepine2007b} for $\zeta=$ 0.1--1.5 in steps of 0.1. The
solid lines delineate the boundary between the dM/sdM/esdM/usdM
classes ($\zeta=$ 0.825, 0.500, and 0.200, respectively); the single-star
spectral standards for the sdM (diamonds), esdM (triangles), and usdM
(squares) classes are also shown \citep{Lepine2007b}. For clarity, only pairs
whose error bars, in both CaH and TiO5 of both pairs, are smaller than the
median error are plotted.  Most of the observed pairs are dMs, i.e.,
part of the thin disk with roughly solar metallicity. This is not
surprising for a bright sample located within $\sim$200--300~pc of the
Sun as the local neighborhood is largely dominated by thin disk stars
\citep{Bochanski2010}. 

Most of the binary pairs lie parallel to the iso-$\zeta$ lines,
within the error bars. The inset in Figure~\ref{Fig: met0} shows the
distribution of $\Delta\zeta$, which is centered around zero but has
substantial scatter. Quantitatively, the median and median absolute
deviation (m.a.d.) are -0.005 and 0.066, respectively, indicating
that for the vast majority of the observed pairs, $\zeta$ is a
correctly infers a common metallicity for the two stars in each pair.

To test whether the observed $\Delta\zeta$ distribution was merely the intrinsic scatter in
$\Delta\zeta$, we randomly selected 113 pairs of disk stars ($\zeta=$
0.825--1.100) in the similar brightness range as the {\slowpokes}
sample ($r<20$) and with high proper motions ($\mu\ge 40$~{\masyr})
from the SDSS spectroscopic catalog \citep{West2011} and calculated
the $\Delta\zeta$ distribution. There were 8030 stars in the DR7
spectroscopic sample that met these criteria; 113 pairs were randomly
selected from this sample and their $\Delta\zeta$ distribution
calculated. We performed this simulation $10^5$ times; the resultant
distribution is plotted in red in the inset of
Figure~\ref{Fig: met0}. The simulated $\Delta\zeta$ distribution is
less centrally peaked, and much broader, compared to our observed
distribution. The Kolmogorov-Smirnov test \citep{Press1992} finds
a probability of 3.3$\times 10^{-25}$ that the two distributions were drawn
from the same parent population. We conclude that the metallicity of
components of {\slowpokes} binaries observed in this program are more
similar to each other than that of two randomly paired thin disk stars. As 
components of a binary system are expected to have formed of the same
material, this further serves to confirm the physical
association of the pairs. Furthermore, it strengthens the argument
that $\zeta$ is a reliable  proxy for relative metallicity.

However, Figure~\ref{Fig: met0} also demonstrates some deficiencies in
the definition of $\zeta$. First, $\Delta\zeta$ is more than three
m.a.d. away from zero for $\sim$18\% of the pairs, versus the $\sim$5\%
expected for a normal distribution. They are discrepant
especially at large values of TiO5 and CaH2$+$CaH3, i.e., for
higher-mass M dwarfs, perhaps suggesting a break in the $\zeta({\rm
  TiO5, CaH2+CaH3})$ relation. Large errors in this regime further
complicate the issue, as the discrepant $\zeta$ values could result
from the difficulty of measuring the shallower TiO5, CaH2, and
CaH3 band heads in late-K and early-M dwarfs. It is also evident 
how the iso-$\zeta$ contours converge at the higher masses as pointed out out by
\citetalias{Lepine2007b}. On the other hand, the discrepancy persists
for the higher-mass pairs with smaller error bars, as can be clearly
seen in Figure~\ref{Fig: met0}. Second, and  perhaps  more importantly, our
measured $\zeta$ values increase and become super-solar (i.e., $\zeta
> 1$) for the higher mass stars. This is inconsistent with the expectation:
given the apparent magnitude constraints ($r\sim$ 15--17), the higher
mass stars in this sample can be expected to be farther away and,
hence, at larger Galactic heights given most of the SDSS sight lines
are at at high Galactic latitudes \citep{Ivezic2008b}. Stars at high
latitudes are, on average, older; and consequently, if anything, they
might be expected to have lower metallicities \citep{West2008a}. Yet,
the $\zeta$-index yields the opposite. This result necessitates a
redefinition of $\zeta$. With a spectroscopic sample of 113 visually
resolved binaries, we are in an unique position to modify the
definition of $\zeta$.

Given the lack of subdwarf pairs in our sample, any
recalibration of $\zeta$ would be systematically biased to high
metallicity. Hence, we conducted a search for companions around the
subdwarfs ($\zeta < 0.825$) in the SDSS DR7 spectroscopic catalog
\citep{West2011} with extant SDSS spectra. We have identified a sample
of ten pairs with low values of $\zeta$; and they 
are shown as open circles in Figure~\ref{Fig: met0}. The full
sample that was identified in the search will be presented in a future
paper (Dhital et. al., {\it in prep.}). One of the added pairs is at
the sdM/esdM boundary while the other nine are at the dM/sdM
boundary. While small this sample provides an invaluable constraint
in the low-metallicity regime.

\begin{center}
\begin{deluxetable}{lccccc}
  \tablecaption{Coefficients, $a_N$, for Eq (\ref{Eq: TiO5})}
  \tablecolumns{8}
  \tabletypesize{\scriptsize}
  \tablewidth{0pt}
  \tablehead{
    \colhead{Coefficients} &
    \colhead{\citetalias{Lepine2007b}} &
    \colhead{This study}
}
\startdata

a$_0$ & $-$ 0.050 & $-$ 0.047 \\
a$_1$ & $-$ 0.118 & $-$ 0.127 \\
a$_2$ & $+$ 0.670 & $+$ 0.694 \\
a$_3$ & $-$ 0.164 & $-$ 0.183 \\
a$_4$ & \nodata   & $-$ 0.005
\enddata
\label{Tab: zeta}
\end{deluxetable}
\end{center}

\citetalias{Lepine2007b} defined $\zeta$ as:
\begin{equation}\label{Eq: zeta}
\zeta = \frac{1- {\rm TiO5}}{1 - [{\rm TiO5}]_{Z_\odot}},
\end{equation}
where [TiO5]$_{\rm Z_{\odot}}$ is a third-order polynomial of (CaH2+CaH3):
\begin{equation}\label{Eq: TiO5}
[{\rm TiO5}]_{\rm Z_{\odot}} = \sum_N a_N ({\rm CaH2+CaH3})^N,
\end{equation}
and where the coefficients, a$_{\rm N}$, are tabulated in Table~\ref{Tab:
  zeta} and were obtained as a single fit to the TiO5 index as a function of
CaH2+CaH3 index for kinematically-selected sample of thin disk stars. 

We can recalibrate $\zeta$
by varying the functional form of [TiO5]$_{\rm Z_{\odot}}$ in
Eq.~(\ref{Eq: TiO5}) such that the scatter in the $\Delta\zeta$
distribution is minimized and distributed around zero. As noted
earlier, the \citetalias{Lepine2007b} definition, to first-order, is 
a robust measure of relative metallicity; and a recalibration need only
be a perturbation about that definition. Moreover, as the definition
was based on the distribution of (TiO5, CaH2+CaH3) of disk stars, it
is a good starting point for the recalibration. So we only chose to
explore the coefficient values within $\pm$0.03 of the
\citet{Lepine2007b} values in steps of $\Delta=$ 0.001. We have
introduced a fourth-order term Eq. (\ref{Eq: TiO5}), with an initial
guess of zero, based on the observed deviation of higher-mass pairs
from the iso-$\zeta$ lines. The best fit values for the coefficients
were found by minimizing $\chi^2$, where $\Delta\zeta=0$ was assumed
to be the model. All dM/sdM pairs, except for the ones with the large error
bars, were considered for the fit. 

\begin{figure*}[tbh!]
  \begin{centering}
    \includegraphics[width=1\linewidth]{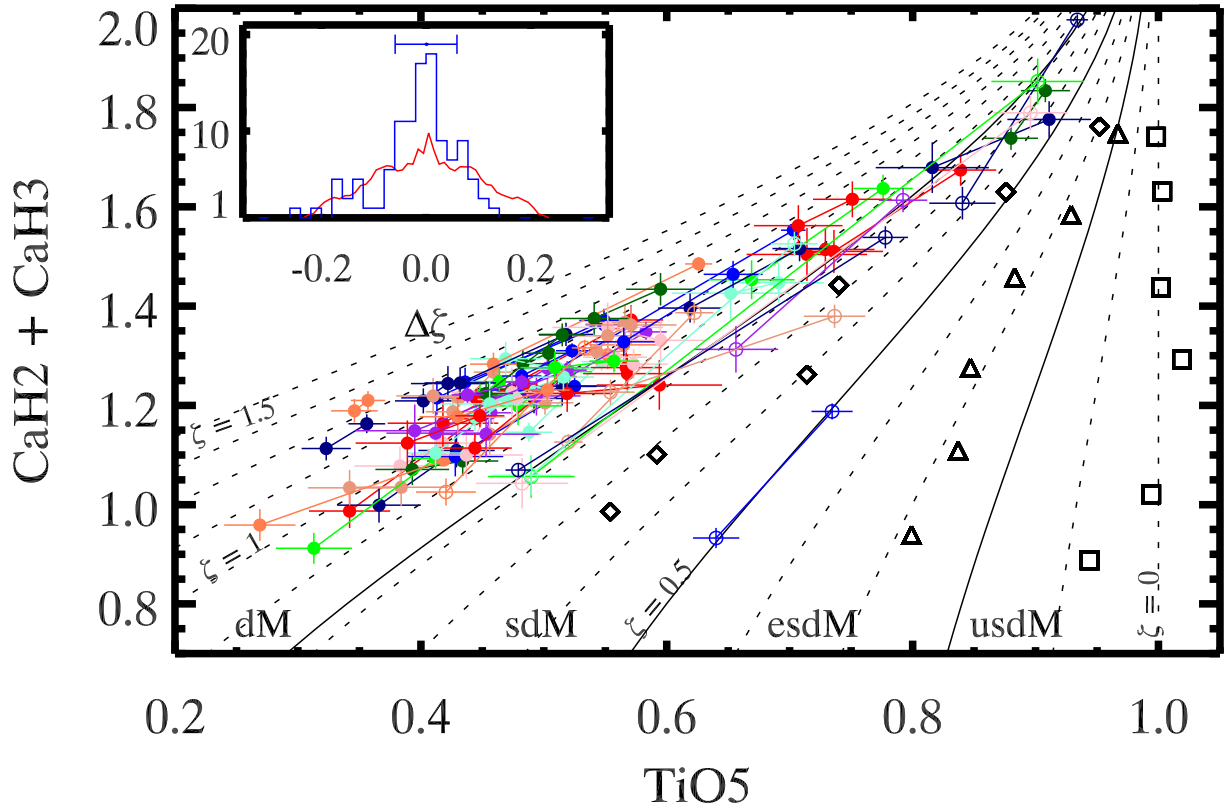}
    \caption{\protect\small
      Same as Figure~\ref{Fig: met0} but with the redefined $\zeta$,
      which was done by adding a fourth-order term in Eq. (\ref{Eq:
        TiO5}) and refitting the coefficients by perturbing about the
      \citet{Lepine2007b} values. The new definition of $\zeta$
      minimizes the scatter in the $\Delta\zeta$ distribution and
      yields lower metallicity for stars at larger Galactic heights, which
      are expected to be older and have a lower metallicity. The
      $\Delta\zeta$ distribution is now also much more centrally
      peaked compared to random pairings of unassociated stars
      (compare to inset of Figure~\ref{Fig: met0}.}
    \label{Fig: met1}
  \end{centering}
\end{figure*}

Figure~\ref{Fig: met1} shows the new iso-$\zeta$ contours, with the
coefficients tabulated in Table~\ref{Tab: zeta}. The contours look
significantly different despite small changes in the coefficients,
reflecting the very sensitive dependence of $\zeta$ on its independent
variables. The differences can be summarized as:

\begin{enumerate}
  \item The scatter in the $\Delta\zeta$ is smaller, with the
    m.a.d. decreasing from 0.060 to 0.044. There are fewer outliers as
    well, with the values converging towards  $\Delta\zeta =0$. As the
    median error in  $\Delta\zeta$ is bigger than the scatter in
    $\Delta\zeta$, decreasing the scatter further is not possible
    unless higher S/N data are obtained. The $\Delta\zeta$
    distribution is much more centrally peaked compared to randomly
    paired field stars.
  \item With only a few pairs at the higher-masses (large TiO5), $\zeta$
    yields more consistent values between components of the higher-mass
    dMs than in the original \citetalias{Lepine2007b} definition. This
    congruence is important if we are to have the same 
    metallicity proxy for all low-mass stars and is likely to improve as
    more pairs are added to that locus. It is especially reassuring to
    get the same value of $\zeta$ for the components of pairs with
    large differences in mass, CaH2+CaH3, and TiO5; the previous
    definition such pairs were especially discrepant as the primary
    was assigned a artificially super-solar $\zeta$. With the new
    definition, the higher-mass dMs instead have slightly sub-solar
    $\zeta$, which is what is expected for stars at larger Galactic heights.
  \item The new iso-$\zeta$ contours are less crowded at the higher-mass
    regime, allowing for a more robust determination of
    metallicity. In fact, the iso-$\zeta$ do not converge before
    (TiO5, CaH2+CaH3) $=$ (1,2), hence, expanding the regime for
    sdMs/esdMs/usdMs. The metallicity classes also become more
    sensitive to TiO5 relative to CaH2+CaH3. However, with the new
    contours for $\zeta$, the previously defined standards for the
    metallicity classes are assessed to be more metal-poor and no
    longer are in the same class. While this argues
    for a definition of new standards, we advise against such a
    revision until there are more subdwarf binaries to more robustly
    calibrate the contours in that regime.
\end{enumerate}

In general, the new definition of $\zeta$ better fits the observed
sample of visual binaries as well as resolving outstanding issues at
the high-mass end. However, due to a paucity of subdwarf pairs, it
leaves the low-metallicity regime rather unconstrained. $\zeta$ looks
to be a good proxy for metallicity, and future observations of
subdwarf pairs should calibrate it for all low-mass stars. Studies
that are using subdwarf binaries (Dhital et al., {\it in prep.}) and
F/G--K/M binaries (Bochanski et al. {\it in prep.}; Lepine et al. {\it
  in prep.})  are already underway and are part of a larger effort to
measure the absolute metallicity of low-mass stars. 

\subsection{{\ha} Activity}\label{Sec: age-act}
As M dwarf photospheres are too cold to excite electrons into the $n=2$
level, any observed {\ha} feature, either in absorption or in emission,
is chromospheric in origin. The weakest chromospheres
will exhibit no {\ha}; as the activity levels increase, {\ha} will be
observed in absorption with the line filling in and eventually going
into emission for the most active M dwarfs \citep{Stauffer1986,
  Cram1987, Walkowicz2009}. As {\ha} is in absorption for both
inactive and moderately active M dwarfs, {\ha} emission has
traditionally been used as the tracer of chromospheric activity and is
biased towards the most active M dwarfs \citep[e.g.,][]{West2011}.

In our sample, 11 of the 113 pairs showed clear evidence of {\ha}
emission in both components while three pairs had only one component with
{\ha} in emission. In addition, 33 pairs showed {\ha} absorption in
both components and 22 pairs in one component; the remaining 47 were
classified as inactive. The large fraction (39\%) of stars 
with {\ha} absorption is in accordance with the nearby M dwarfs in the Palomar/MSU Nearby Star Survey
Spectroscopic Survey \citep{Gizis2002a}. Overall, for 91 of the 113 (81\%)
pairs in our sample, both components of a pair showed similar
levels of activity---in emission, absorption, or the lack or activity. 
For the pairs with both components having {\ha} emission, we converted
the equivalent width in {\ha} to {\lha}---the ratio of {\ha}
luminosity to the bolometric luminosity that is independent of
spectral type---following \citet{Walkowicz2004} and \citet{West2004}; all pairs
had comparable levels of {\lha} within the error bars except for the
two where the active primary was identified as a candidate SB and had
an inactive secondary. The tidal forces due to the tight companion has
presumably enhanced the activity of the primary \citep{Shkolnik2010,
  Silvestri2006}.

As {\ha} activity depends strongly on mass \citep{West2008a}, one way to
compare the intrinsic variability in activity levels is by only
looking at pairs with components of similar masses. All sixteen
pairs with {\ha} in emission or absorption in both components and with
similar colors ({$\Delta$(\rz)} $<$ 0.2) had {\ha} equivalent widths within
130\% of each other. Compared to the 200--300\% difference in {\ha}
activity exhibited by M dwarfs over time \citep{Bell2011}, the
components of the similar-mass binary pairs in our sample exhibit a
much smaller difference in {\ha} activity. This is consistent with the
expectation that stars of similar ages and masses have comparable
activity levels, presumably because they experience similar spin-down
rates.

\section{Conclusions}\label{Sec: conclusions}
We have carried out a spectroscopic follow-up study of 113 ultra-wide,
low-mass CPM binary systems from the {\slowpokes} catalog
\citepalias{Dhital2010} using the GOLDCAM spectrograph on the KPNO
2.1~m telescope. We measured the radial velocities of each component
by cross-correlating them with appropriate standards and used them to
assess the fidelity of pairs in the {\slowpokes} catalog. 95 of the
113 (84\%) of the pairs have the same radial velocity within
3~$\sigma$. At least five of the pairs with discrepant radial
velocities are candidate SBs, which would explain the
difference. There may be additional spectroscopic companions
undetected in our low-resolution spectra. \citet{Law2010} found that
45\% of the {\slowpokes} systems are either hierarchical triples or
quadruples. Either high-resolution spectroscopy or imaging would be
needed to identify the close companions and to further quantify the
incidence of higher-order systems in wide binaries.

We examined the {\ha} activity in our observed sample. The components
of binary pairs exhibited overwhelmingly comparable levels of {\ha}
activity. Moreover, the {$\Delta$\ha} of the pairs with similar {\rz}
colors and two active components, while large, was several times
smaller than the variation seen in single M dwarfs. Our results corroborate that
low-mass stars of the same mass should spin-down at the similar rates
over time. However, larger samples of active are needed to confirm
this finding and to constrain the rate of this spin-down.

We tested the \citetalias{Lepine2007b} $\zeta$-index and found
that, to first-order, it is a robust measure of relative
metallicity. The value of $\zeta$ for the two components in each binary
system match within the error bars for most pairs, indicating a common
metallicity as expected. However, we find a systematic bias for
the higher-mass M dwarfs such that $\zeta$ overestimates the
metallicity. Assuming all of the pairs are physically associated
systems and have the same metallicity, we have redefined
$\zeta$. While the shift is small, it better represents
iso-metallicity lines in the high-metallicity regime and represents an
incremental step towards defining an absolute metallicity scale for
low-mass dwarfs. Planned further observations should extend the
calibration of $\zeta$ as well as map it to an absolute metallicity
scale in the near future.

\acknowledgements
We would like to acknowledge Kevin R. Covey and the anonymous referee
for valuable discussion and comments on the paper.
SD, AAW, and KGS; AAW; and JJB acknowledge funding support through
NSF grants AST--0909463, AST--1109273, and AST--0544588, respectively.
SD would like to thank NOAO for supporting travel to and 
from Kitt Peak as part of the Dissertation Support Program. APM
acknowledges Boston University's Research Internship in Science \&
Engineering Program for funding her summer internship at Boston
University.  The authors would like to thank Daryl Willmarth (KPNO)
and William Sherry (NOAO) for their help during the 2009A observing run.  

Funding for the SDSS and SDSS-II has been provided by the Alfred
P. Sloan Foundation, the Participating Institutions, the National
Science Foundation, the US Department of Energy, the National
Aeronautics and Space Administration, the Japanese Monbukagakusho, the
Max Planck Society, and the Higher Education Funding Council for
England. The SDSS Web site is \url{http://www.sdss.org}. 

The SDSS is managed by the Astrophysical Research Consortium for the
Participating Institutions. The Participating Institutions are the
American Museum of Natural History, Astrophysical Institute Potsdam,
University of Basel, University of Cambridge, Case Western Reserve
University, University of Chicago, Drexel University, Fermilab, the
Institute for Advanced Study, the Japan Participation Group, Johns
Hopkins University, the Joint Institute for Nuclear Astrophysics, the
Kavli Institute for Particle Astrophysics and Cosmology, the Korean
Scientist Group, the Chinese Academy of Sciences (LAMOST), Los Alamos
National Laboratory, the Max-Planck-Institute for Astronomy (MPIA),
the Max-Planck-Institute for Astrophysics (MPA), New Mexico State
University, Ohio State University, University of Pittsburgh,
University of Portsmouth, Princeton University, the United States
Naval Observatory, and the University of Washington. 

We acknowledge use of the ADS bibliographic service.

Facilities: \facility{KPNO}

\bibliography{ads}
\end{document}